\documentstyle[twocolumn,epsf]{article}
\makeatletter
\renewcommand{\section}{%
\@startsection{section}{1}{\z@}
{8truept}{4truept}{\normalsize\bf}}
\renewcommand{\subsection}{%
\@startsection{subsection}{1}{\z@}
{8truept}{4truept}{\normalsize\bf}}
\def\fnum@figure{Fig.\hskip.5em\thefigure}
\long\def\@makecaption#1#2{
  \setbox\@tempboxa\hbox{{\bf #1}\quad #2}%
  \ifdim \wd\@tempboxa >\hsize \unhbox\@tempboxa\par
  \else \hbox to\hsize{\hfil\box\@tempboxa\hfil}
  \fi}
\makeatother
\begin{document}
\pagestyle{empty}
\title{\Large\bf Large Thermopower in a Layered Oxide NaCo$_2$O$_4$}
\author{\large I. Terasaki$^{1,2,*}$, Y. Sasago$^3$ and K. Uchinokura$^3$\\
{\normalsize $^1$Department of Applied Physics, Waseda University,
Tokyo 169-8555, Japan}\\
{\normalsize $^2$Precursory Research for Embryonic Science and
Technology, Japan Science Technology Corporation}\\
{\normalsize $^3$Department of Applied Physics,
The University of Tokyo, Tokyo 113-8656, Japan}}
\date{}
\maketitle
\thispagestyle{empty}

\section*{Abstract}
A transition-metal oxide NaCo$_2$O$_4$ is a layered oxide
in which CoO$_2$ and Na alternately stack along the $c$ axis. 
Recently we have found that this compound 
shows large thermopower with low resistivity,
which is comparable to those of Bi$_2$Te$_3$.
The negative transverse magnetoresistance 
and the strongly temperature-dependent Hall coefficient
suggest that electron correlation dominates  
the conduction mechanism in NaCo$_2$O$_4$.

\section{Introduction}
Transition-metal oxides are known as a large class of materials
where the band width and the carrier density
can be controlled by cation substitutions.
They exhibit various physical properties, {\it e.g.},
ferroelectricity, magnetism and superconductivity.
Since the discovery of the high-temperature superconductors,  
transition-metal oxides have attracted renewed interest,
and a number of new materials and new phenomena
have been discovered in the past decade.

The search for new thermoelectric (TE) materials
has also come to a new stage where ternary or quarternary 
compounds are extensively examined \cite{Mahan}.  
A filled Skutterudite is a prime example of the newly 
discovered TE materials \cite{Sales}.  
Very recently we discovered that a layered oxide
NaCo$_2$O$_4$ shows large thermopower and low resistivity,
whose power factor is comparable to that for Bi$_2$Te$_3$ \cite{Tera}.
This strongly suggests that NaCo$_2$O$_4$ is a possible
candidate for a TE transition-metal oxide.

NaCo$_2$O$_4$ is an old material, 
which was synthesized in 1970's \cite{JH},
even though the metallic conduction has been reported recently \cite{Tanaka}.
Figure 1 shows the schematic view of the oxygen network 
of the CoO$_2$ block in NaCo$_2$O$_4$.
Edge-shared distorted octahedra (and Co in the center of them)
form a triangular lattice.
Na cations and CoO$_2$ blocks alternately stack 
along the $c$ axis to make a layered structure.  

A characteristic feature is that the mobility 
for NaCo$_2$O$_4$ is ten times smaller than that for Bi$_2$Te$_3$.
In other words, the carrier density for NaCo$_2$O$_4$ is
ten times larger, implying that small carrier density
is not an origin for the large thermopower.
Hence there should be another mechanism to cause large thermopower, 
and the elucidation of the conduction mechanism in NaCo$_2$O$_4$
may give a new principle to design TE materials.  
Here we report the measurements of the various transport 
parameters of NaCo$_2$O$_4$ single crystals,
and discuss the anomalous conduction mechanism quantitatively.

\begin{figure}[t]
\centerline{\epsfxsize=7cm 
\epsfbox{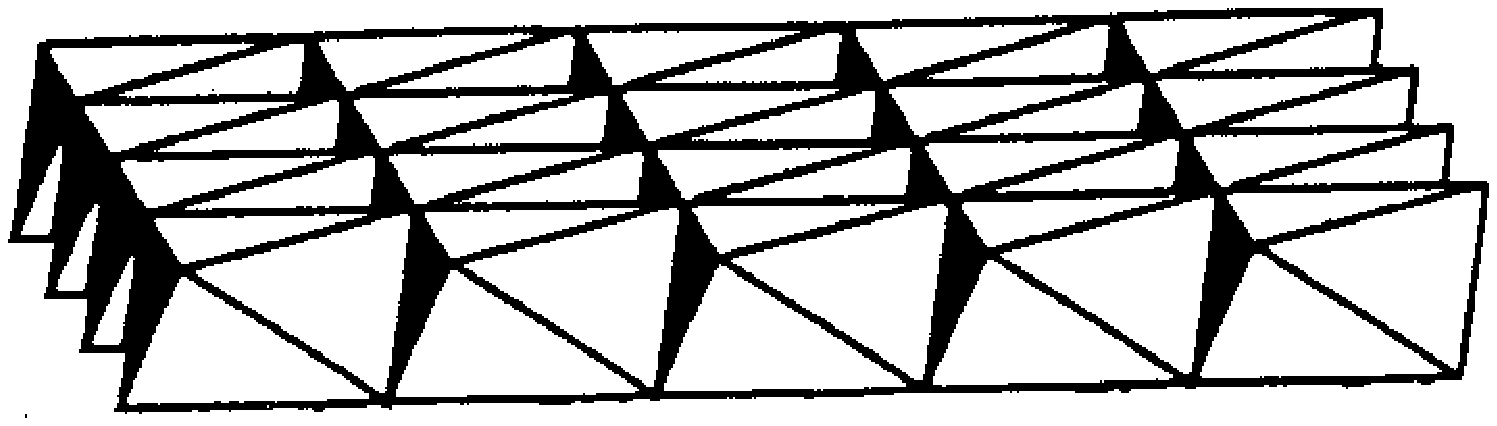}
}
\caption{The oxygen network of CoO$_2$ block in NaCo$_2$O$_4$.}
\end{figure}

\section{Experimental}
Single crystals of NaCo$_2$O$_4$ were prepared by 
a NaCl-flux technique with Al$_2$O$_3$ crucibles.
The as-grown crystals were washed in water to remove the NaCl flux.
The crystals were thin along the $c$ axis with a typical dimension of
1.5$\times$1.5$\times$0.02 mm$^3$.

Transport parameters were measured only along the in-plane direction,
since the crystals were very thin along the out-of-plane direction.
Thermopower was measured in the configuration
where one edge of the sample was pasted on a sapphire plate 
with the other pasted on a sheet heater.
Temperature ($T$) was monitored by two diode thermometers
attached on the edges.
The contribution of copper leads was carefully subtracted.
Transverse magnetoresistance and the Hall coefficient
were measured in sweeping field ($H$) 
from 0 to 8 T using an ac-bridge nano-ohm-meter (Linear Research LR201).

\begin{figure}[t]
\centerline{\epsfxsize=7cm 
\epsfbox{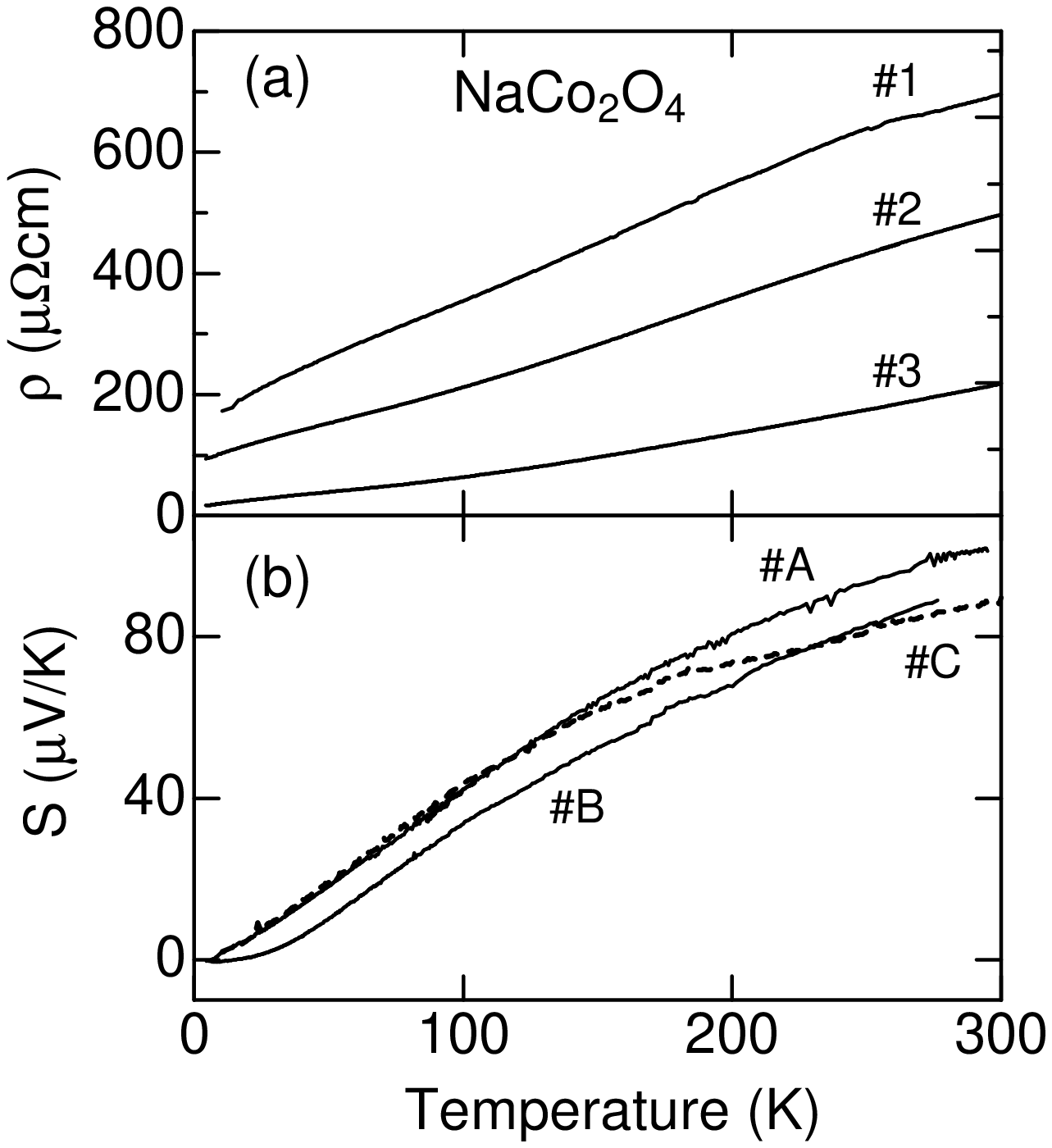}
}
\caption{(a) Resistivity and (b) thermopower 
of NaCo$_2$O$_4$ single crystals.
Different samples are denoted by \#.}
\vspace*{1cm}
\centerline{\epsfxsize=7cm 
\epsfbox{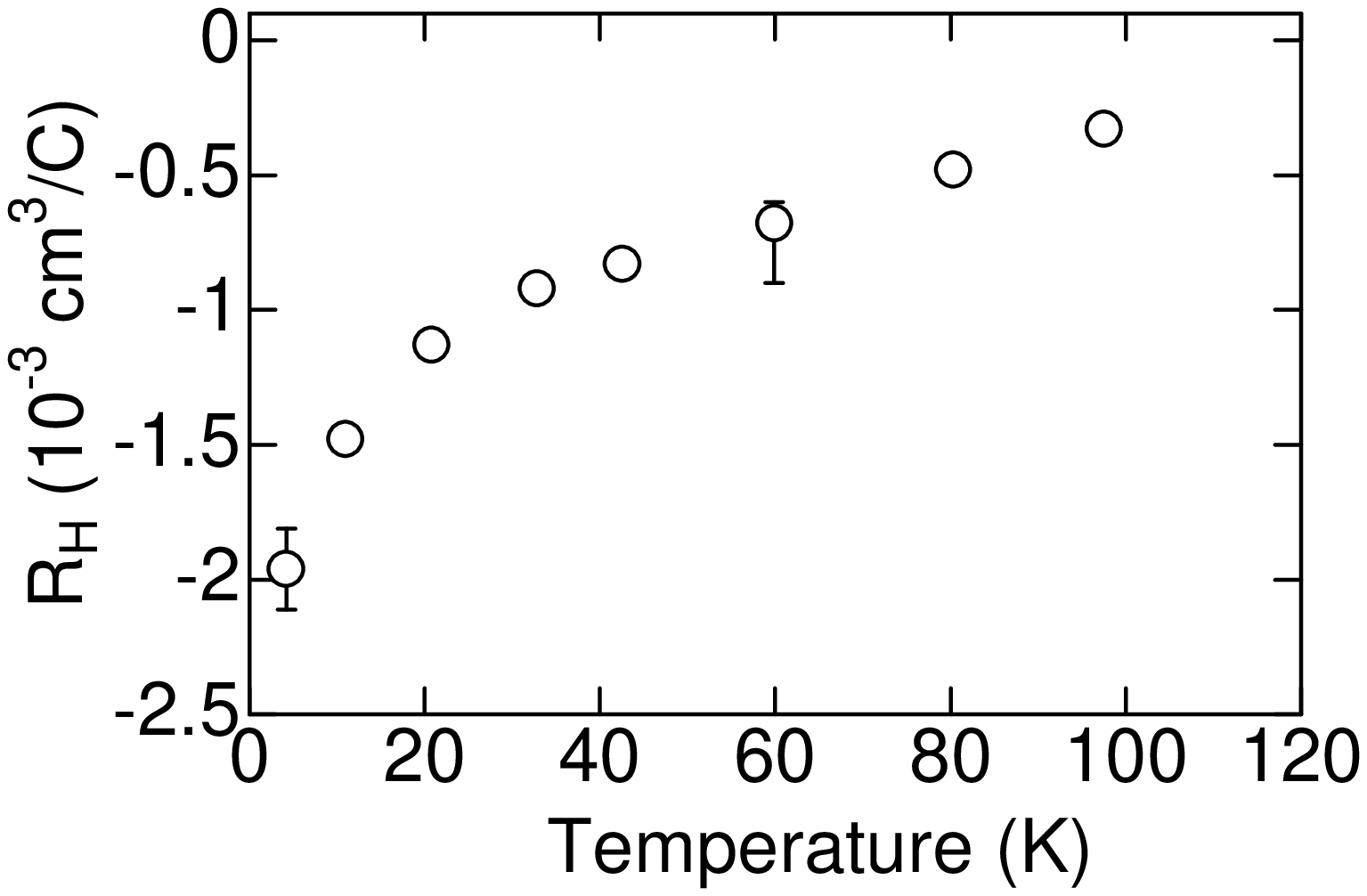}
}
\caption{Hall coefficient of a NaCo$_2$O$_4$ single crystal.
Current is applied along the in-plane direction,
and magnetic field is parallel to the $c$ axis.}
\end{figure}

\section{Results}
Figure 2(a) shows the resistivity ($\rho$) of 
NaCo$_2$O$_4$ single crystals.
The magnitude of resistivity ranges from 200 to 600 $\mu\Omega$cm 
at 300 K, which is possibly due to the uncontrollable disorder
or nonstoichiometry of the Na content.
We observed that $\rho$ of polycrystalline Na$_{1.1+x}$Co$_2$O$_4$,
monotonically increases with $x$ \cite{Itoh}.
In contrast to $\rho$,
the thermopower ($S$) is less sensitive 
to the crystal quality [Fig.~2(b)].  
It should be noted that the magnitude of $S$ (100 $\mu$V/K at 300 K)
is one-order-of-magnitude larger than conventional metals.
As is often observed in conventional metals,
$S$ is roughly proportional to $T$,
which is consistent with the fact that $\rho$ 
exhibits metallic conduction from room temperature down to 1.5 K.

\begin{figure}[t]
\centerline{\epsfxsize=7cm 
\epsfbox{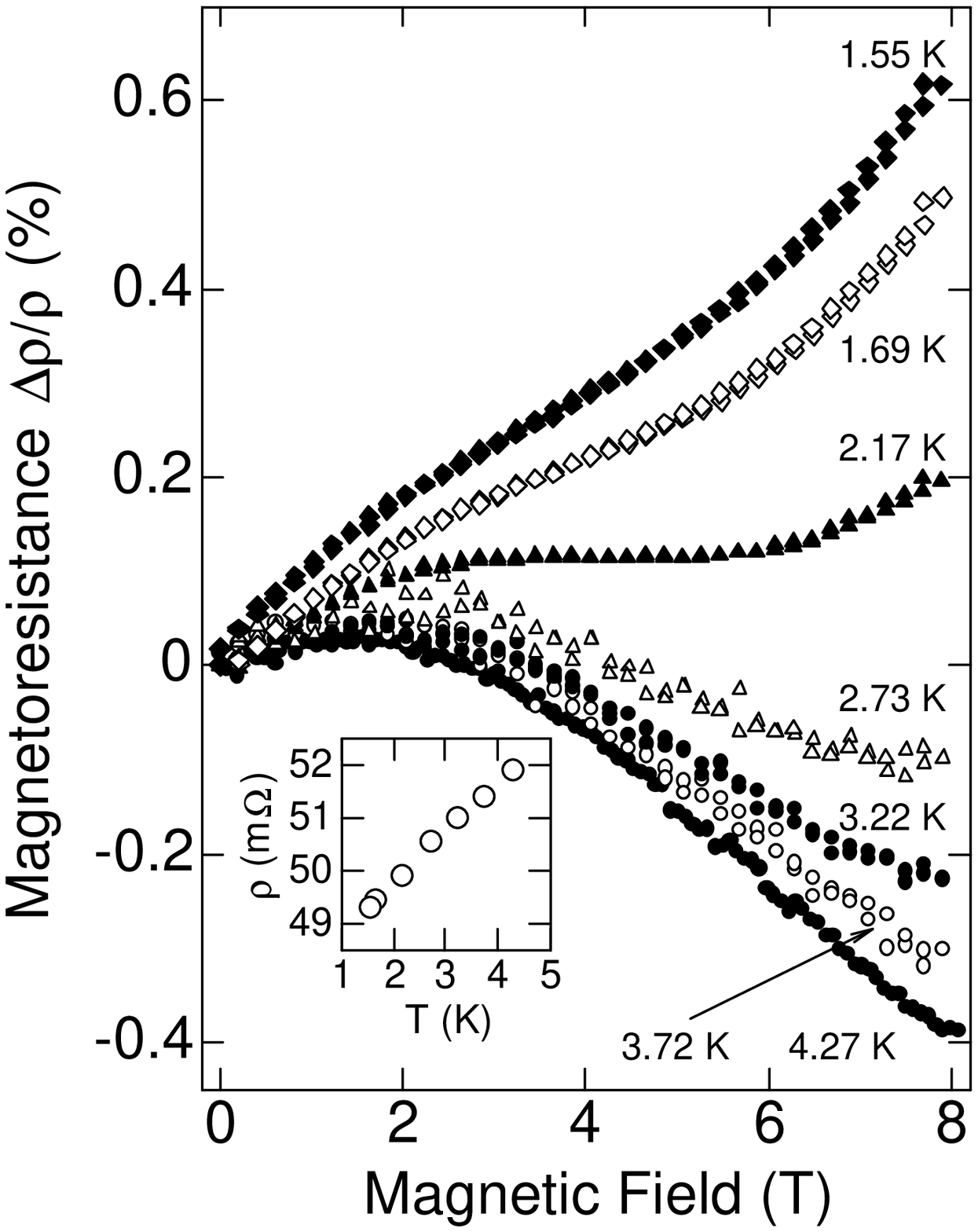}
}
\caption{Transverse magnetoresistance 
of a NaCo$_2$O$_4$ single crystal.
Current is applied along the in-plane direction,
and magnetic field is parallel to the $c$ axis.
The inset shows the zero-field resistance of the sample.}
\end{figure}

The Hall coefficient ($R_H$) is plotted as a function of $T$ in Fig.~3.
The magnitude of $R_H$ at 4.2 K is 2$\times$10$^{-3}$ cm$^3$/C,
corresopnding to the carrier density of 10$^{21}$ cm$^{-3}$.
This is consistent with the fact that NaCo$_2$O$_4$ has ten times larger 
carrier density than Bi$_2$Te$_3$.
Note that $R_H$ shows the opposite sign to $S$.
This clearly indicates that the electronic states cannot 
be understood by a simple parabolic band picture.
Another feature is the remarkable $T$ dependence.
In conventional metals, $1/R_H$ is proportional to 
the carrier density (or density of states at the Fermi energy),
and is independent of $T$ in the lowest-order approximation.

Figure 4 shows the transverse magnetoresistance (MR),
where negative MR is observed at 4.2 K.
As $T$ is lowered, positive MR (roughly proportional to $H^2$)
overlaps with the negative MR.
A possible origin for the negative MR is weak localization
or scattering by spin fluctuation.
Considering that the metallic conduction continues down to 1.55 K
(see the inset of Fig.~4),
the former is unlikely to occur.
Magnetic measurements have shown that spin fluctuation 
may exist in NaCo$_2$O$_4$.
The susceptibility of NaCo$_2$O$_4$ shows Curie-Weiss-like
$T$ dependence \cite{Tanaka},
and the Knight shifts of the Na and Co sites show
different $T$ dependence \cite{Chiba}.
We further note that $\rho$ depends on $T$ even below 4.2 K,
which indicates that the carriers are not scattered by ordinary phonons.

\section{Discussion}
\subsection{Electric conduction mechanism}
As is seen in the previous section,
the large thermopower, the $T$-dependent Hall coefficient,
and the negative magnetoresistance
are difficult to explain by a conventional picture 
based on the band theory and the electron-phonon scattering.
All these results strongly suggest that,
like high-$T_c$ superconductors,
the strong electron correlation is important 
in the electric conduction in NaCo$_2$O$_4$.

We can expect that the strong correlation enhances $|S|$ 
under certain conditions.
Since the diffusive part of $S$ corresponds to the 
transport entropy \cite{KU},
larger electronic specific heat can give larger $S$. 
Thus $S$ would be enhanced if the carriers
could couple with some outside entropy such as 
optical phonon, spin fluctuation, or orbital fluctuation.
A similar scenario has been applied to heavy fermions
or valence-fluctuation systems,
some of which show large $S$ \cite{HF}.
The $T$ dependence of $\rho$ and $S$ of NaCo$_2$O$_4$ is,
at least qualitatively, consistent with 
the theories of 2D metals with spin fluctuation by
Moriya, Takahashi and Ueda \cite{Moriya} 
and by Miyake and Narikiyo \cite{Miyake}.

\subsection{Effect of layered structure}
It should be noted that the half of the Na sites are randomly vacant,
that is, the Na layer is highly disordered.
Thus the mean free path (MFP) of phonons will be as short as
the lattice spacing,
which means that the lattice thermal conductivity is minimized.
In fact, the thermal conductivity of polycrystalline NaCo$_2$O$_4$
is as small as that of Bi$_2$Te$_3$,
and it is hardly affected by cation substitutions \cite{Yakabe}.

On the other hand, electric conduction is determined by the 
CoO$_2$ block that includes little disorder.
In low-dimensional correlated metals, 
the carriers are often confined in the conducting region,
and are hardly affected by outside disorder.
This is known as the ``confinement'' \cite{Anderson,Clarke}. 
As a result, the electric conduction remains considerably good,  
together with the minimized lattice thermal conductivity.
In other words, MFP of carriers can be much longer
than MFP of phonons in NaCo$_2$O$_4$.

We propose that strongly correlated layered conductors
are promising as new TE materials,
in the sense that lattice thermal current and electric current can 
flow in different paths in the crystal.
In this context we may call them a new type of
`electron crystals and phonon glasses' \cite{Sales}. 

\section{Summary}
In summary,
we prepared single crystals of metallic layered 
transition-metal oxide NaCo$_2$O$_4$,
and measured the thermoelectric power,
the Hall coefficient, and the magnetoresistance.
All the measured quantities are unconventional:
(i) the thermoelectric power is unusually large 
(100 $\mu$V/K at 300 K),
(ii) the Hall coefficient strongly depends on temperature, 
and (iii) the magnetoresistance is negative at 4.2 K.
These results indicate that the conduction mechanism in
NaCo$_2$O$_4$ is not explained by a conventional band-picture
and electron-phonon scattering.

\section*{Acknowledgements}
The authors would like to thank M. Takano, S. Nakamura,
K. Fukuda and K. Kohn for fruitful discussions.
They also appreciate H. Yakabe,
K. Nakamura, K. Fujita and K. Kikuchi for collaboration.

\end{document}